\documentclass{llncs}
\usepackage{amsopn}
\usepackage{color}
\usepackage[square,numbers,comma,sort&compress]{natbib}
\usepackage{enumerate}
\usepackage{colortbl}
\usepackage{booktabs}
\usepackage{iba-algo}
\usepackage{ifpdf}
\usepackage{graphics}
\usepackage{epsfig}
\usepackage{subfigure}
\usepackage{booktabs}
\usepackage{mathrsfs}  
\usepackage{setspace}
\usepackage[normalem]{ulem}
\usepackage{amsmath}
\usepackage{amssymb}
\usepackage{fancyhdr}

\def\ve#1{\mathchoice{\mbox{\boldmath$\displaystyle\bf#1$}}
{\mbox{\boldmath$\textstyle\bf#1$}}
{\mbox{\boldmath$\scriptstyle\bf#1$}}
{\mbox{\boldmath$\scriptscriptstyle\bf#1$}}}

\newcommand\R{\mathbb R}  
  
\newcommand\Po{\mathcal P}

\newcommand\Tr{\mathcal T}

\DeclareMathOperator{\conv}{conv}

\DeclareMathOperator{\supp}{supp}

\DeclareMathOperator{\argmax}{argmax}    
\DeclareMathOperator{\argmin}{argmin}    
\DeclareMathOperator{\vertices}{vert}    

\let\epsilon=\varepsilon
\usepackage{ifthen}
\makeatletter
\newcommand{\DeclareBracket}[3]{
  \newcommand{#1}[2][]{%
  \ifthenelse%
  {\equal{##1}{}}%
  {\left#2##2\right#3}%
  {\csname ##1l\endcsname#2##2\csname ##1r\endcsname#3}}}    
\makeatother
\DeclareBracket\abs||           
\DeclareBracket\norm\|\|        
\DeclareBracket\floor\lfloor\rfloor
\DeclareBracket\ceil\lceil\rceil
\DeclareBracket\set\{\}
\DeclareBracket\paren()
\DeclareBracket\inner\langle\rangle
\DeclareBracket\fractional\{\}

\definecolor{purple}{RGB}{160,32,240}       
\definecolor{darkyellow}{RGB}{190,190,0}       

\renewcommand{\thetable}{\Roman{table}}

\newtheorem{algorithm}{Algorithm}

\title{Optimality of the Neighbor Joining Algorithm and Faces of the Balanced Minimum Evolution Polytope}
\titlerunning{Optimality of the NJ Algorithm \& Faces of the BME Polytope}

\author{David C. Haws\inst{1}, Terrell L. Hodge\inst{2},  and Ruriko Yoshida\inst{1}}
\authorrunning{Haws, Hodge, Yoshida}
\institute{University of Kentucky, Lexington, KY, 40502-00227 \and
Western Michigan University, Kalamazoo, MI, 49008-5248}

\begin{document}

\maketitle
\thispagestyle{fancy}


\begin{abstract}
 
Balanced minimum evolution (BME) is a statistically consistent distance-based
method to reconstruct a phylogenetic tree from an alignment of molecular data.
In 2000, Pauplin showed that the BME method is equivalent  to optimizing  a
linear functional over the BME polytope,  the convex hull of the BME vectors
obtained from Pauplin's formula applied to all binary trees. The BME method is
related to the Neighbor Joining (NJ) algorithm,  now known to be a greedy
optimization of the BME principle.   Further, the NJ and BME algorithms have
been studied previously to understand when the NJ Algorithm returns a BME tree
for small numbers of taxa. In this paper we aim to elucidate the structure of
the BME polytope and strengthen knowledge of the connection between the BME
method and NJ Algorithm. We first prove that any subtree-prune-regraft move
from a binary tree to another binary tree corresponds to an edge of the BME
polytope. Moreover, we describe an entire family of faces parametrized by
disjoint clades. We show that these {\em clade-faces} are smaller dimensional
BME polytopes themselves. Finally, we show that for any order of joining nodes
to form a tree, there exists an associated distance matrix (i.e., dissimilarity
map)  for which the NJ Algorithm returns the BME tree. More strongly, we show
that the BME cone and every NJ cone associated to a tree $T$ have an
intersection of positive measure.

\end{abstract}
\begin{flushleft}
{\bf Keywords:} Phylogentic Analysis, Balanced Minimum Evolution, Neighbor
Joining, Polyhedral Geometry, Combinatorics.
\end{flushleft}

\section{Introduction}

Current efforts to reconstruct the tree of life for different organisms demand
the inference of phylogenies from thousands of DNA sequences (see
http://tolweb.org/tree/ \cite{zr} and \cite{Ciccarelli:2006ys} for more
details).  Large scale projects include the investigation of the tree of life
for flies, by researchers at North Carolina State University \\
(\url{http://www.inhs.illinois.edu/research/FLYTREE/}), the tree of life for
fungi, at Duke University (\url{http://aftol.org}/), and at the University of
Kentucky, the tree of life for the insect order Hymenoptera
(\url{http://www.hymatol.org/}). 

The most established approach to tree reconstruction is the maximum likelihood
(ML) method. In this method, evolution is described in terms of a
discrete-state continuous-time Markov process on a phylogenetic tree.
Unfortunately, an exhaustive search for the ML phylogenetic tree is
computationally prohibitive for large data sets \cite{Rock2006}.   However, one
can efficiently compute a pairwise distance, a distance between a pair of
leaves, using the ML method.  The pairwise distances can then be used, together
with a distanced-based tree reconstruction method, to recover the phylogenetic
tree that relates the sequences \cite{Felsenstein2003}, albeit at  a loss of
accuracy.  
\pagestyle{plain}
To date, distance-based methods for phylogeny reconstruction have been seen to
be the best hope for accurately building phylogenies on very large sets of taxa
such as the data sets for tree of life for Hymenoptera \cite{Semple2003, BME}.
More precisely, distance-based methods have been shown to be statistically
consistent in all settings  ( such as the long branch attraction) in contrast
with parsimony methods \cite{Felsenstein1978,consis,Gascuel2003,Gascuel2009}.
Distance-based methods also have a huge speed advantage over parsimony and
likelihood methods, and hence enable the reconstruction of trees on greater
numbers of taxa.

In 2002, Desper and Gascuel introduced a balanced minimum evolution (BME)
principle, based on a branch length estimation scheme of Pauplin
\cite{Pauplin}.  The guiding principle of minimum evolution tree reconstruction
methods is to return a tree whose total length (sum of branch lengths) is
minimal, given an  input dissimilarity map.  The BME method is a special case
of these  distance-based methods  wherein branch lengths are estimated  by a
weighted least-squares method (in terms of the input dissimilarity map and the
tree in question)  that puts more emphasis on shorter distances than longer
ones. Each labeled tree topology gives rise to a vector, called herein {\em the
BME vector,}  which is obtained from Pauplin's formula.

Implementing, exploring, and better understanding the BME method have been
focal points of several recent works.   The software {\tt FastME},  developed
by Desper and Gascuel,  heuristically optimizes the BME principle using
nearest-neighbor interchanges (NNI) \cite{fastme}. In simulations, {\tt FastME}
gives superior trees compared to other distance-based methods, including one of
biologists' most popular distance-based methods, the Neighbor Joining (NJ)
Algorithm, developed by Saitou and Nei \cite{Saitou1987}.  In 2000, Pauplin
showed that the BME method is equivalent to optimizing a linear function, the
dissimilarity map,  over the BME representations of binary trees, given by the
BME vectors \cite{Pauplin}. Eickmeyer et. al. defined the {\it $n^{th}$ BME
polytope} as the convex hull of the BME vectors  for all binary trees on a
fixed number $n$ of taxa.  Hence the BME method is equivalent to optimizing a
linear function, namely, the input dissimilarity map, over a BME polytope.  In
2010, Matsen and Cueto \cite{Matsen2010} studied how the BME method works when
the addition of an extra taxon to a data set alters the structure of the
optimal phylogenetic tree.  They characterized  the behavior of the BME
phylogenetics on such data sets, using the BME polytopes and the {\it BME
cones}, i.e., the normal cones of the BME polytope.

Eickmeyer et. al. studied the BME polytopes computationally,  for unrooted
phylogenetic trees with eight or fewer taxa.  In addition to this computational
study of the BME polytopes, they showed the following general lemma:
\begin{lemma}[Lemma 3.1 in \cite{kord2009}]
For any number of taxa $n,$ the vertices of the $n^{th}$ BME polytope are
exactly the BME vectors of all unrooted binary trees with $n$ leaves.  The BME
vector of the star phylogeny lies in the interior of the BME polytope, and all
other BME vectors  lie on the boundary of the BME polytope.
\end{lemma}

In particular, Eickmeyer et. al. studied edges of the BME polytopes
computationally.
\begin{figure}[!ht]
\begin{center}
 \includegraphics[scale=0.7]{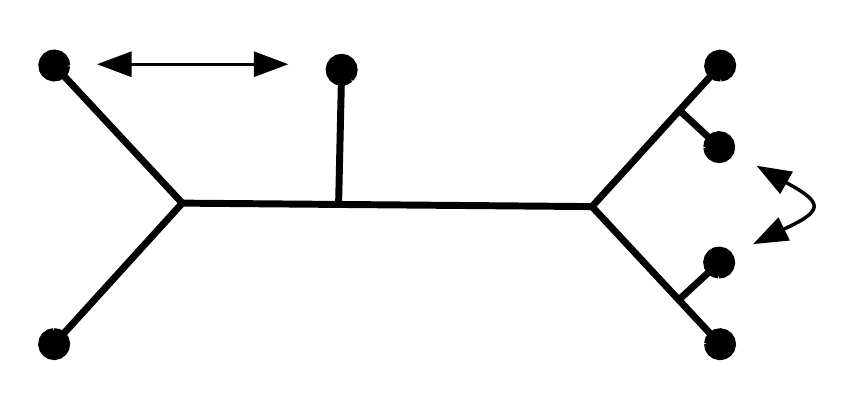}
\end{center}
\caption{The non-edges on the $n^{th}$ BME polytope for $n = 7$.  Two trees
will form a non-edge if and only if they are trees that have three cherries,
and differ by the pair of leaf exchanges shown in the figure.}
\label{fig:nonedge}
\end{figure}
They found that the edge graph of the $n^{th}$ BME polytope is the complete
graph $T_2$ with the same number ($\leq 6$) of leaves, there is a dissimilarity
map for which $T_1$ and $T_2$ are (the only) co-optimal BME trees. However, for
$n=7$, the BME polytope has one combinatorial type of non-edge, i.e., the BME
vectors of two bifurcating trees with seven leaves and three cherries (two
leaves adjacent to the unique internal node in the tree)  fail to be joined by
an edge if and only if  their trees  are related by two leaf exchanges as
depicted in Figure \ref{fig:nonedge}. This completely characterizes the
non-edges for $n=7$. 
  
Characterizing the edges of the $n^{th}$ BME polytope for $n > 7$ remains an
open problem that motivated this work. Understanding the structure of the BME
polytope through its edges and faces may help with the development of  new
optimization strategies to find an optimal BME tree. For example, one such
approach could entail employing an edge-walking method over the edges of the
BME polytope,  since the BME method is a linear programming problem over the
BME polytope.  However,  until now, not much was known about the  faces of the
BME polytopes besides vertices (which are trivial to characterize). 

This paper makes contributions towards understanding both edges and
higher-dimensional faces of the BME polytope.  First, we prove that any
subtree-prune-regraft (SPR) move from a binary tree to another binary tree
corresponds to an edge of the BME polytope.  This implies that any NNI move
from a binary tree to another binary tree corresponds to an edge of the BME
polytope.  Consequently, the method implemented in the software {\tt FastME} is
an edge-walking method over the edges of the BME polytope using NNI moves.
Moreover, we define and describe an entire family of faces of the BME polytope
that are parametrized by disjoint clades.  We show that these {\em clade-faces}
are smaller dimensional BME polytopes themselves. 

 The study of related geometric structures, the BME cones, further clarifies
 the nature of the link between phylogenetic tree reconstruction using the BME
 criterion and using the Neighbor Joining (NJ) Algorithm.  In 2006, Gascuel and
 Steel showed that the NJ Algorithm, one of the most popular phylogenetic tree
 reconstruction algorithms, is a greedy algorithm for finding the BME tree
 associated to a dissimilarity map \citep{Steel2006}.  The Neighbor Joining
 Algorithm relies on a particular criterion for iteratively selecting cherries;
 details on cherry-picking and the NJ Algorithm are recalled later in the
 paper.  In 2008, based on the fact that the selection criterion for
 cherry-picking is linear in the dissimilarity map \cite{Bryant2005}, Eickmeyer
 et. al. showed that the NJ Algorithm will pick cherries to merge in a
 particular order and output a particular tree topology $T$ if and only if the
 pairwise distances satisfy a system of linear inequalities, whose solution set
 forms a polyhedral cone in  ${\R}^{n \choose 2}$ \cite{kord2009}.  They
 defined such a cone as an {\em NJ cone}.  In general, the sequence of cherries
 chosen by the NJ Algorithm is not unique, hence multiple dissimilarity maps
 will be assigned by the NJ Algorithm to a single fixed tree topology $T.$  The
 set of all dissimilarity maps for which  the NJ Algorithm returns a fixed tree
 topology $T$ is a union of NJ cones, however this union is not convex in
 general. Eickmeyer et. al. \cite{kord2009} characterized those dissimilarity
 maps for which the NJ Algorithm  returns the BME tree, by comparing the NJ
 cones with the BME cones, for eight or fewer taxa. 

Yet, before this paper, it was unclear whether, given a tree topology $T$ with
an arbitrary number of taxa, and any particular order of picking cherries
allowed by the NJ Algorithm, there existed a dissimilarity map such that the NJ
Algorithm would return the BME tree $T$.  We prove this in fact is so, despite
the fact that greedy algorithms do not generally construct the globally optimal
structure for the condition which they locally optimize.  Interpreted in terms
of phylogenetics, this is particularly important, as it shows that there is no
order of picking cherries for which the NJ Algorithm will fail to return the
BME tree. Geometrically this means that for any NJ cone associated with  the
tree topology $T$ and a particular choice of cherry-picking order, there exists
a non-empty intersection with the BME cone associated with $T.$ Consequently,
given any tree topology $T$, there exists a dissimilarity map such that NJ and
BME both return the tree topology $T$. More strongly, we show that the BME cone
and every NJ cone associated to a tree $T$ have an intersection of positive
measure.

This paper is organized as follows: Definitions and notation are covered in
Section \ref{sec:def}.  Subsection \ref{subsec:cladeface}  treats clade-faces
of the BME polytope and contains a useful proposition concerning objective
criteria for greedy linear optimization.  Section \ref{sec:sprbme} contains the
proof that two trees adjacent by an SPR move form an edge of the BME polytope.
In Section \ref{sec:cfa} we present the Cherry Forcing Algorithm and show that
it also provides proof for the existence of clade-faces.  Finally, using the
Cherry Forcing Algorithm, in Section \ref{sec:njbme} we prove that every NJ
cone associated with a tree $T$ has a non-empty intersection  of positive
measure with the BME cone associated with $T$. That is,  given a tree $T$ and a
sequence of cherries chosen by the NJ Algorithm, there is a dissimilarity map
such that NJ and BME return $T$. We finish with a discussion in Section
\ref{sec:disc}.

\section{Notation, Definitions and Further Preliminaries}
\label{sec:def}
\subsection{Phylogenetic $X$-trees, Cherries, and Clades}
Let $X$ be a set of leaves, which we also may call taxa;  when $|X| = n,$ we
will often conveniently identify $X$ with $\{1, 2, \ldots n\}.$  A {\em
dissimilarity map} (or {\em distance matrix}) is a function $d: X\times X \to
\mathbb{R}$ with $d(x,x) = 0$ and $d(x,y) = d(y,x)$ for all $x,y\in X.$ It is
convenient to represent a dissimilarity map by a vector $\ve d \in \R^{n
\choose 2}.$ In general, we index entries of any $\ve c \in \R^{n \choose 2}$
by pairs $\{i,j\} \subset X $ with $i < j$ in lexicographic order, i.e. $\ve c
= (c_{12},c_{13},\ldots,c_{1n},c_{23},\ldots, c_{2,n},\ldots,c_{n-1,n}) \in
\R^{n \choose 2}$.  We may also index a set of vectors in $\R^{n \choose 2}$ by
superscript when necessary,  e.g., $\ve c^k\in \R^{n\choose 2}$ with $ijth$
coordinate $c^k_{ij}.$ Define $\ve e_{ij} \in \R^{n \choose 2}$ to be the
vector with $1$ at the $ij$th entry and $0$ else.  Let  $\R^{n \choose 2}_+ =
\{\ve x \in \R^{n \choose 2}\,|\,  x_{ij} \geq 0 \mbox{ for all }1 \leq i < j
\leq n\}$.

Mathematically, a {\em tree} is an undirected graph in which any two vertices
are connected by exactly one simple path; the number of edges incident to any
vertex (i.e., node) $x$ of the tree is the degree $deg(x)$ of $x.$ If  the
graph consists of more than a single vertex, a node $x$ with $deg(x) = 1$ is
{\em external}, or a {\em leaf}; all other nodes are {\em internal}.  A {\it
phylogenetic} $X$-tree is a tree $T$ with set of leaves  $X$ and all internal
vertices of degree at least three. Those for which the internal vertices are
all of degree three are here called {\it binary $X$-trees} (or just {\em binary
trees}, when the context is clear). For $n = |X| \geq 3$ the binary $X$-trees
are necessarily unrooted trees, and for $n\geq 4,$ correspond in phylogenetics
to unrooted cladograms with no polytomy.  Let $\Tr_n$ be the set of all  binary
trees with $n$ leaves; we will assume throughout $n \geq 3$. Write $E(T)$ for
the set of edges (i.e., branches) of $T\in \Tr_n.$ An edge $e\in E(T)$ is {\em
internal} (resp., {\em external}) if it does not (resp., does) touch a leaf. A
{\it cherry} of $T \in \Tr_n$ is a pair of leaves $\{i,j\}$ such that the path
between them consists of just two (necessarily external) edges.  An {\em
edge-weighting} (or {\em branch length assignment}) $\omega$ of $T$ is a
function $\omega: E(T) \to \mathbb{R}$ with $\omega(e) \geq 0$ for every $e\in
E(T).$  Given an edge weighting $\omega,$ define the {\em total tree length}
$\omega(T):=\sum_{e\in E(T)}\omega(e).$  

An {\em $X$-split} is a partition $A\,|\,B$ of $X$ into two subsets (blocks)
$A,B\subset X.$  Any edge $e  \in E(T)$ of $T \in \Tr_n$ induces an $X$-split
$A_1\,|\,A_2$ by deleting $e$ from $T$ and letting $A_i$ be the subset of
leaves associated to the resulting connected component $C_i,$ $i = 1, 2,$ of
$T.$ Conversely any $X$-split $A_1\, | \, A_2$ corresponds to the edge that
when deleted gives the split. When $e \in E(T)$ is internal, we will call $C_i$
a {\it clade}, and $A_i$ the {\it support} $\supp(C_i)$ of $C_i.$ When the
context is clear, we may identify a clade $C$ with its support $\supp(C).$  By
allowing for the case of choosing no edge $e,$ that is, the trivial $X$-split
$\emptyset\,|\,X,$ we obtain $T$ itself as a clade.  Further, we simply say a
clade $C$ is in $\Tr_n$ (and write $C\in \Tr_n$) if $C$ is a clade for some
tree in $\Tr_n$.  

For example, leaves $\{1,2\}$ define a clade of $T_1$ of Figure
\ref{fig:trees_n4} whereas $\{2,3\}$ is not a clade of $T_1$ since there is no
subgraph containing $\{2,3\}$ attainable by removing an internal edge of $T_1$.
We say two clades $C_1, C_2 \in \Tr_n$ are {\em disjoint} if $\supp(C_1)\cap
\supp(C_2) = \emptyset.$ If $C_1, C_2$ are disjoint clades both contained in a
tree $T \in \Tr_n$, then we define the {\em distance between clades}
$d_T(C_1,C_2)$ as the number of edges between clades $C_1$ and $C_2$ in $T$. 

Finally, given $T\in \Tr_n$, let $\Sigma(T)$ denote the set of $X$-splits
defined by removal of an edge of $T.$ It is well-known \cite{Buneman1971} that
phylogenetic $X$-trees $T_1,T_2 \in \Tr_n$ are determined up to equivalence (as
graphs) exactly when $\Sigma(T_1) = \Sigma(T_2).$ If $\omega: E(T) \to
\mathbb{R}$ is an edge-weighting, and $A\,|\,B$ is a split in $\Sigma(T)$ with
corresponding edge $e\in E(T),$  set $\omega(A\,|\,B) := \omega(e).$  A
distance matrix $\ve c \in \R^{n \choose 2}$ is called an {\em additive metric}
or {\em tree metric} if $\ve c$ is a metric, and there exists a tree $T \in
\Tr_n$ and an edge-weighting $\omega$ on $T$ s.t.
\begin{itemize}
\item[(a)] $\omega(e) > 0$ for all $e\in E(T).$ 
\item[(b)] For every pair of leaves $\{i, \, j\},$ $c_{ij} = \sum_e\omega(e),$
summing over edges $e$ along the path from leaf $i$ to leaf $j$. 
\end{itemize}
\noindent Clearly, given $T\in \Tr_n$ and an edge-weighting $\omega$ on $T,$
setting $D_{T,\omega}(i,j) = \sum_e\omega(e),$ for the sum as in (b) above,
yields a tree metric $D_{T,\omega}\in \R^{n \choose 2}_+$. Given a tree $T \in
\Tr_n$ and any split $A\,|\,B$ in $T$, where $A,B \subseteq \{1,2,\ldots,n\}$,
the {\em split metric} is defined as $D^{A\,|\,B} =  (d_{ij}^{A\,|\,B})$ where
$d_{ij}^{A\,|\,B} = 1$ if $i \neq j$ and $|\{i,j\} \, \cap \, A| = 1$, and
$d_{ij}^{A\,|\,B} = 0$ else. Thus each split $A\,|\,B$ defines a metric
$D^{A\,|\,B}$ from $T$ for which all branch lengths equal to zero, except for
the branch $e$ corresponding to $A\,|\,B.$ For any edge-weighting $\omega$ of
$T,$  the split metrics for $T$ and the natural tree metric $D_{T,\omega}$ are
related as below (see, e.g., \cite{Bandelt:1992uq}): 
\begin{equation}
\label{e:DTomega}
D_{T,\omega} = \sum_{A\,|\,B \in \Sigma(T)} \omega(A\,|\,B)D^{A\,|\,B}.
\end{equation}    

\subsection{Amalgamation of Cherries}

The {\it amalgamation} of $T \in \Tr_n$ by cherry $\{i,j\}$ is the subtree
$\widetilde T$ on $n-1$ leaves   obtained by amalgamating the vertices in
cherry $\{i,j\}$  to their common internal parent node.  In the more formal
mathematical language of relations on the set of leaves $X = \{x_1,\dots,
x_n\},$ the amalgamation of a cherry $\{x_i,x_j\}\in X\times X$ to its common
internal parent node $v_{i,j}$ corresponds to a two-step merge obtained
(without loss of generality) by first merging the nodes $x_i, v_{i,j}$ to a new
(internal) node $v_{i,j}',$ and then merging $x_j, v_{i,j}'$ to a new
(external) node $[x_i,x_j]$, resulting in the new tree $\widetilde{T}$ on the
leaf set $\widetilde{X} = X - \{x_i,x_j\} \cup \{[x_i,x_j]\}.$  

\begin{figure}[!htp]
\begin{center}
\scalebox{0.7}{\includegraphics{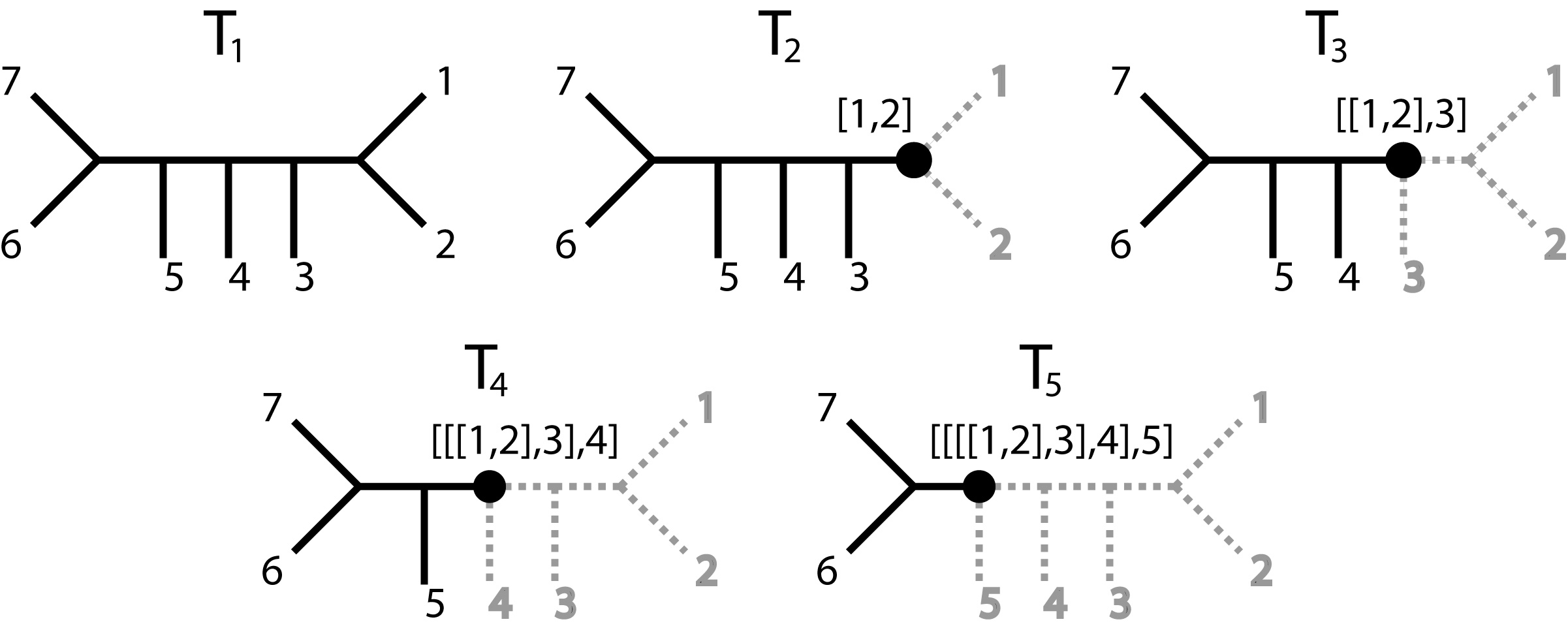}}

\caption{Cherry $\{1,2\}$ of tree $T_1$ is amalgamated yielding tree $T_2,$
where the new leaf is labeled $[1,2]$. Cherry $\{[1,2],3\}$ of $T_2$ is
amalgamated yielding tree $T_3,$ where the new leaf is labeled $[[1,2],3]$.
Similarly, the leaves $\{[[1,2],3],4\}$, $\{[[[1,2],3],4],5\}$ are amalgamated
in trees $T_3$ and $T_4,$ respectively. }
\label{fig:amalgex}
\end{center}
\end{figure}

For example, in Figure \ref{fig:amalgex}, amalgamating cherry $\{1,2\}$ of
$T_1$ gives $T_2$ with the new leaf labeled $[1,2]$. Next, amalgamating cherry
$\{[1,2],3\}$ in $T_2$ produces  $T_3$.  If $T'$ is obtained from $T$ by
successive amalgamations of cherries (including the possibility that no
cherries are amalgamated, so $T' = T$),  then any leaf $i'$ of $T'$ is either
present in $T$ as a single leaf, or the result of the amalgamation of leaves
$i_1,\ldots,i_t$ of $T$. Hence, leaf $i'$ of $T'$ induces the clade $C$ of $T$
with $\supp(C) =  \{i'\},$ in the first case, or $\supp(C) =
\{i_1,\ldots,i_t\},$ in the second case. For example, leaf $[[[1,2],3,4]]$ of
$T_4$ in Figure \ref{fig:amalgex} defines a clade in $T_1$ given by the leaves
$1$,$2$,$3$, and $4$ of $T_1$. We will call the clade of $T$ obtained from any
leaf $i'$ of  $T'$ the \emph{subgraph of $T$ given by $i'$}.

\subsection{Balanced Minimum Evolution: Method, Vectors, and Polytopes}

For a phylogenetic $X$-tree $T \in \mathcal{T}_n$ and a dissimilarity map $\ve
d \in \R^{n \choose 2}_{+},$ there are different biologically relevant methods
to assign branch lengths (i.e., an edge-weighting) to $T$; in this context, the
entry $d_{i,j}$ of $\ve d$  is most often regarded as the distance between any
pair of taxa $i$ and $j$.  The balanced minimum evolution (BME) method employs
a weighted least squares approach for assigning branch lengths $l:E(T) \to
\mathbb{R}^+$ given the dissimilarity map $\ve d$. Defined by Pauplin
\cite{Pauplin}, the definition of the edge-weights $l(e)$ (i.e.,
\cite[Equations (2), (3), equiv., (7), (8)]{BME}) utilizes average distances
between clades whence consequently the $l(e)$ are, moreover, linear in the
input dissimilarity map  $\ve d$ (e.g., see Equation (1) in \cite{BME}).
However, for the BME method, the calculation of the total tree length $l(T) =
\sum_{e\in E(T)}l(e)$ can be easily stated and quickly computed without
resorting to computing individual branch lengths $l(e)$, by the means we now
describe. For any pair $\{i,j\}$ of leaves of $T,$ define $y^T_{ij} := \#\{ $
edges between leaves $i$ and $j\,\}$, the topological distance between $i$ and
$j$. Set $w_{ij}^T := 2^{1-y^T_{ij}}$, so $\ve w^T := (w_{12}^T, w_{13}^T,
\ldots, w_{n-1,n}^T) \in \R^{n \choose 2}$ is a vector depending only on the
topology of $T.$   Pauplin's \cite{Pauplin} formula for the {\em balanced tree
length estimation} (or {\em estimated BME length}) $l(T)$ is given by 
\begin{equation}
 l(T) = \sum_{i,j: i<j} w^T_{i,j}d_{i,j} = \ve w^T \cdot \ve d.
\label{eq:pauplin}
\end{equation}
When necessary for clarity, we will also indicate the dependence on $\ve d$ of
the estimated BME length $\ve w^T \cdot \ve d$ by writing $l(T,\ve d).$  Since
$\ve w^T$ depends only on the topology of $T$, but determines $l(T,\ve d)$
given any input dissimilarity map $\ve d\in \R^{n \choose 2}_+$, we call $\ve
w^T$ the {\it BME vector for $T$.} 

In \cite{Steel2004}, the authors in fact defined terms  $w^T_{i,j}$ for any
phylogenetic $X$-tree $T$ (not necessarily binary) in terms of certain cyclic
permutations  of (``circular orderings'') of $X$ that respect the structure of
$T$ as measured through its set of splits $\Sigma(T).$ In the case of
edge-weighted binary $X$-trees, one recovers the expression for $w^T_{i,j}$ in
the BME vector and Pauplin's formula.  \cite{Steel2004} used this perspective
to establish the consistency of the balanced tree length estimation. That is,
if $T\in \Tr_n$ has branch lengths $\omega$ and one takes $\ve d =
D_{T,\omega}$ in Equation \ref{eq:pauplin}, one obtains $l(T) = \omega(T).$ 
 
The {\em BME method} for phylogenetic tree reconstruction (or {\em BME
principle})  can be succinctly stated:  find a $T \in \Tr_n$ such that Equation
\ref{eq:pauplin} is minimized, given the dissimilarity map $\ve d  \in \R^{n
\choose 2}$. 
 
Note that one can efficiently compute the input for the BME method, i.e.,
pairwise distances $d(i,j),$ from any given sequence alignment using the
maximum likelihood estimators (MLEs) under an evolutionary model. The BME
method for tree reconstruction was shown to be consistent in \cite{BME}.

We recall some necessary definition from polyhedral geometry
\cite{Schrijver1986Theory-of-linea}. The \emph{convex hull} of $\{\ve
a_1,\ldots,\ve a_m\} \subset \R^n$ is defined as \begin{equation*}   \conv
\{\ve a_1,\ldots,\ve a_m\} := \left\{\, \ve x \in \R^n \mid \ve x =
\sum_{i=1}^m \lambda_i \ve a_i, \, \sum_{i=1}^m \lambda_i = 1, \, \lambda_i
\geq 0 \, \right\}.  \end{equation*} A \emph{polytope} $\Po$ is the convex hull
of finitely many points. We say $F \subseteq P$ is a {\em face} of the polytope
$\Po$ if there exists a vector $\ve c$ such that $F = \argmax_{\ve x \in \Po}
\ve c \cdot \ve x$. Every face $F$ of $\Po$ is also a polytope. If the
dimension of $\Po$ is $d$, a face $F$ is a \emph{facet} if it is of dimension
$d-1$. A face is an \emph{edge} if it is of dimension two.  Denote the vertex
set of a polytope $\Po$ by $\vertices(\Po),$ where a vertex of a
$d$-dimensional polytope is the intersection point of $d$ or more edges, faces
or facets.

With the background on BME above in hand, we now recall the definition of the
central object of study in this paper, the BME polytope, as it arises from the
BME vectors. 

\begin{definition}[BME polytope]
The {\it balanced minimum evolution} (BME) polytope $\Po_n$ on $n$ leaves is defined as

\begin{equation*}
    \Po_n := \conv \left\{\, \ve w^T \mid T \in \Tr_n\,\right\}\notag .
\end{equation*}

\end{definition}

If $F$ is a face of $\Po_n$, then its vertex set is given by  $\vertices(F)=
\{\ve w^{T_1},\ldots, \ve w^{T_m}\,|\,\ve w^{T_i}\in F\},$ which we may
identify with the set of trees $\{T_1,\ldots,T_m\}$.  With this definition we
can see that minimizing Equation \ref{eq:pauplin} is equivalent to minimizing
the linear objective $\ve d \in \R^{n \choose 2}$ over $\Po_n$.  Using Day's
results it can be shown that choosing a minimizing tree for \eqref{eq:pauplin}
from among the $(2n-5)!!$ unrooted binary trees is an NP-hard problem
\cite{Day1987,kord2009}. Thus it is NP-hard to optimize linearly over $\Po_n$
\cite{kord2009}.

\begin{figure}[!htp]
\begin{center}
\subfigure[For $|X|=4$, there are the $3$ binary trees and the star-shaped tree.]{
\includegraphics[height=1.9cm]{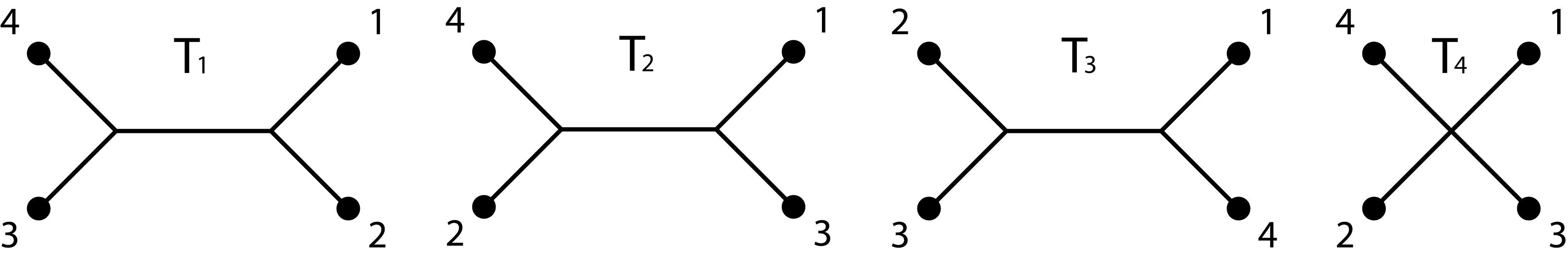}
\label{fig:trees_n4}}
\subfigure[BME polytope on four taxa.]{
\scalebox{0.6}{\ifpdf
    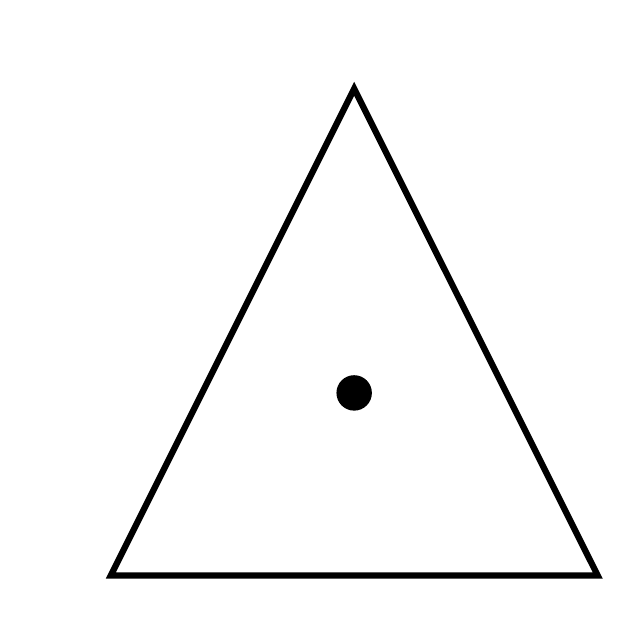
    \else
    \input{BME_n4_ex.pstex_t}
    \fi}
\label{fig:BME_n4_ex}}
\caption{All $X$-trees on four taxa and the BME polytope $\Po_4$}
\end{center}
\end{figure}

\begin{example}[\cite{kord2009}]
For $n=4$, there are the $3$ binary trees and the star-shaped tree as in Figure
\ref{fig:trees_n4}. For this case the BME polytope is the convex hull of the
vectors:
\begin{align*}
& \ve w^{T_1} = \left(\frac{1}{2},\frac{1}{4},\frac{1}{4},\frac{1}{4},\frac{1}{4},\frac{1}{2}\right), \;\; \ve w^{T_2} = \left(\frac{1}{4},\frac{1}{2},\frac{1}{4},\frac{1}{4},\frac{1}{2},\frac{1}{4}\right),\\
&\ve w^{T_3} = \left(\frac{1}{4},\frac{1}{4},\frac{1}{2},\frac{1}{2},\frac{1}{4},\frac{1}{4}\right), \;\;  \ve w^{T_4} = \left(\frac{1}{3},\frac{1}{3},\frac{1}{3},\frac{1}{3},\frac{1}{3},\frac{1}{3}\right)
\end{align*}
Thus the BME polytope $\Po_n$ for $n = 4$ is a triangle in $\R^6$. Note that
the star-shaped tree is in the interior of $\Po_n$.
\end{example}

\begin{remark}[\cite{kord2009}]
The $n^{th}$ BME polytope $\Po_n$ lies in $\R^{{n \choose 2}}$ and has
dimension ${n
  \choose 2} - n$.
\end{remark}

\subsection{Clade Faces}
\label{subsec:cladeface}

With minor modifications of the proof of BME consistency in \cite{BME} we will
show that any collection of disjoint clades defines a face of the BME polytope.
This will also be proved independently and constructively in Section
\ref{sec:njbme} using the Cherry Forcing Algorithm.
\begin{lemma}
Let $C_1,\ldots,C_p \in \Tr_n$ be a pairwise disjoint collection of clades.
There exists a $\ve c \in \R^{n \choose 2}$ such that $\underset{T \in
\Tr_n}{\argmax}\, \ve w^T \cdot \ve c  = \{\, T \in \Tr_n \mid C_1,\ldots,C_p
\in T \,\}$.
\label{lem:bmeconstcladeface}
\end{lemma}
See the appendix for a proof of Lemma \ref{lem:bmeconstcladeface}.

Lemma \ref{lem:bmeconstcladeface} proves that every disjoint set of clades
corresponds to a face of $\Po_n$ which we define as follows: Given a set of
disjoint clades $\{C_1,\ldots,C_p\,|\, C_i \in \Tr_n, \,\forall \,1\leq i\leq
p\}$, we define a {\it clade-face} of the BME polytope $\Po_n$ by
$F_{C_1,\ldots,C_p} := \{\, T \in \Tr_n \mid C_1,\ldots,C_p \in T\,\}$.
Moreover, the face $F_{C_1,\ldots,C_p}$ is the image of an affine
transformation of the BME polytope $\Po_l$, where $l := n - \sum_{i=1}^p (|C_i|
-1 )$. This follows since every tree in $F_{C_1,\ldots,C_p}$ can be constructed
by starting with a binary tree on $l$ leaves and attaching the clades
$\{C_1,\ldots, C_p\}$ to $p$ of the $l$ leaves. 
    
Looking ahead to Section \ref{sec:sprbme}, one can  see that the three trees
corresponding to a nearest neighbor interchange (explained therein) form a
clade-face, as an immediate consequence of Lemma \ref{lem:bmeconstcladeface}.
This suggests that  NNI and SPR moves yield edges of the BME polytope $\Po_n$,
but this fact will  require additional proof.  Leading into this, we provide a
proposition which holds for any polytope in general, and will be key to our
further arguments. Roughly speaking it states that if the entries of $\ve c_1$
are significantly larger than the entries of $\ve c_2$, then when linearly
optimizing $\ve c_1 + \ve c_2$ over a polytope $\Po$ then $\ve c_1$ must be
maximized foremost. 

\begin{proposition}
Let $\Po \subseteq \R^m$ be a polytope and $\ve c_1, \ve c_2 \in \R^m$. 
    If 
\begin{equation}
\min_{\substack{\ve x, \ve y \in \vertices (\Po)\\ \ve x \in \argmax_{\ve z \in \Po}{\ve c_1 \cdot \ve z} \\ \ve c_1 \cdot \ve x \neq \ve c_1 \cdot \ve y}}  \ve c_1 \cdot \ve x - \ve c_1 \cdot \ve y  > \max_{\ve x, \ve y \in \vertices (\Po)} | \ve c_2 \cdot \ve x - \ve c_2 \cdot \ve y |
\label{eq:gengreed}
\end{equation}
then
\begin{equation*}
\underset{\ve z \in \vertices (\Po)}{\argmax}\, (\ve c_1 + \ve c_2) \cdot \ve z \subseteq \underset{\ve z \in \vertices (\Po)}{\argmax}\, \ve c_1 \cdot \ve z.
\end{equation*}
\label{prop:gengreed}
\end{proposition}
 See the appendix for a proof of Proposition \ref{prop:gengreed}.

Consider two clade-faces $F_{C_1,\ldots,C_k}$ and $F_{C'_1,\ldots,C'_{k'}}$
where $|F_{C_1,\ldots,C_k}| > 1$.  Then, $F_{C'_1,\ldots,C'_{k'}} \subseteq
F_{C_1,\ldots,C_k}$ if and only if for every $1 \leq i \leq k$, clade $C_i$ is
contained in, or equal to, clade $C'_j$ for some $1 \leq j \leq k'$. If
$F_{C_1,\ldots,C_k} = \{T\}$, then $F_{C'_1,\ldots,C'_{k'}} \subseteq
F_{C_1,\ldots,C_k}$ if $F_{C'_1,\ldots,C'_{k'}} = \{T\}$.  We note that this
induces a partial order on the clade-faces of $\Tr_n$, and gives a lattice if
one also considers $\Po_n$ and the empty set as  clade-faces. 

\section{SPR Adjacency Implies BME Adjacency}
\label{sec:sprbme}
\begin{figure}[ht]
\begin{center}
\subfigure[]{
\scalebox{1.0}{\includegraphics{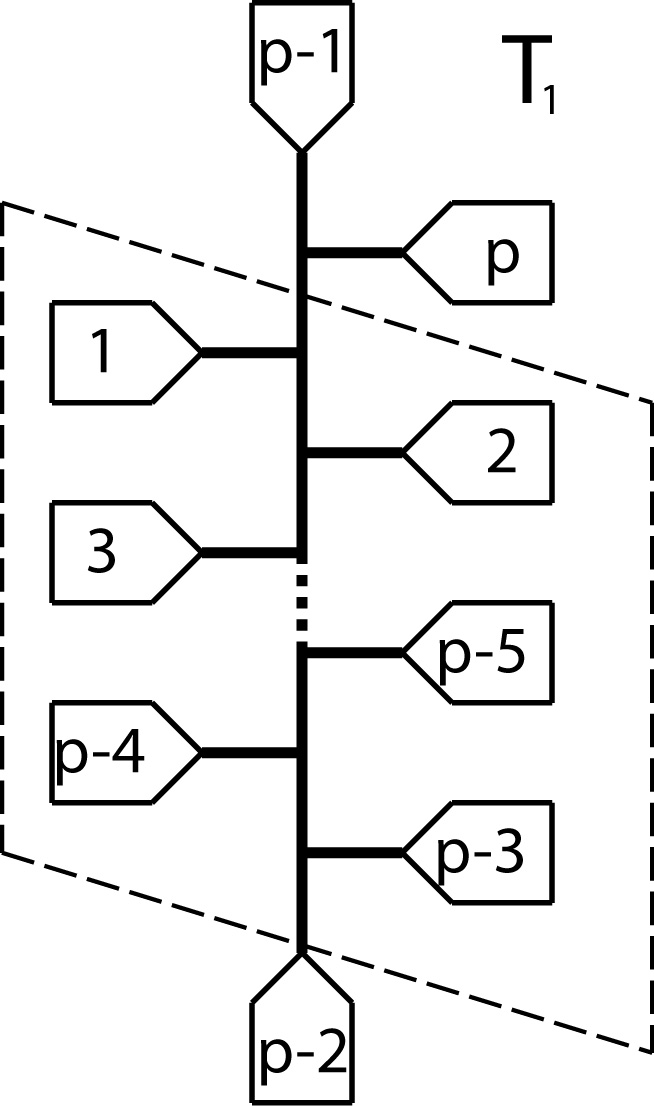}}
\label{fig:sprbase_1}
}
$\phantom{AAAA}$
\subfigure[]{
\scalebox{1.0}{\includegraphics{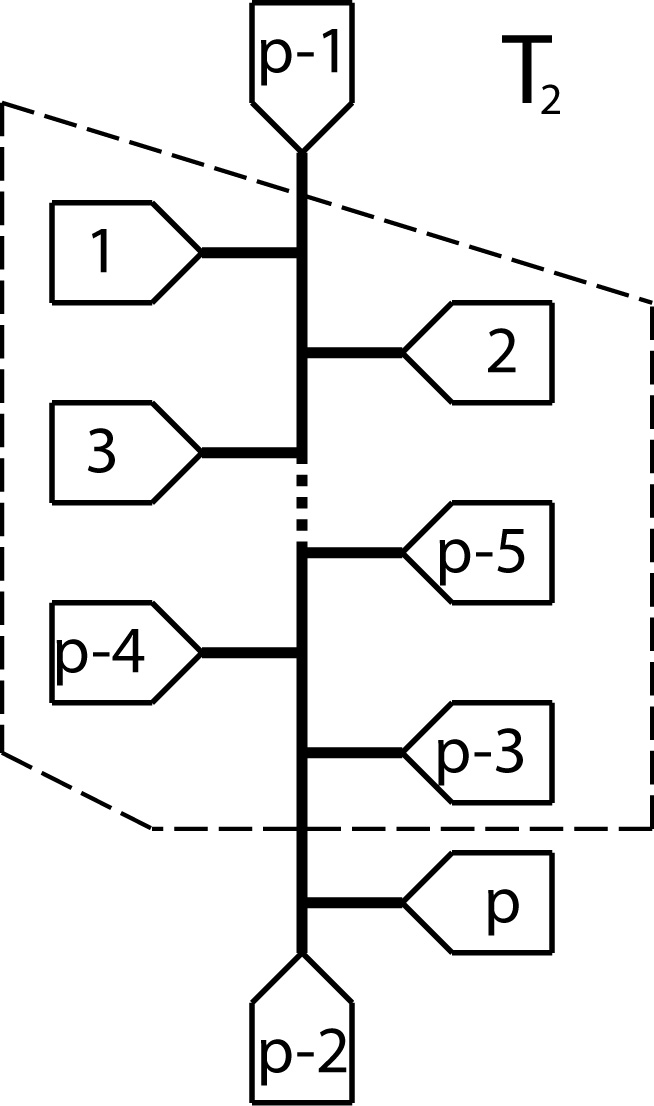}}
\label{fig:sprbase_2}
}
\caption{Here $1,\ldots,p$ are subgraphs (clades). Two trees, $T_1$ (a) and
$T_2$ (b), adjacent by an SPR move where subgraph $p$ and its connecting edge
is the subgraph pruned from $T_1$ and regrafted between subgraph $(p-2)$ and
its internal node.}
\label{fig:sprbase}
\end{center}
\end{figure}

A {\it subtree-prune-regraft} (SPR) move on a tree $T \in \Tr_n$ is determined
by choosing a clade of $C$ of $T$, pruning it from $T$, and amalgamating the
two internal edges originally connecting $C$ to $T$ to one edge. Finally an
internal edge of $T$ is chosen, a node is inserted, and $C$ is attached to this
node. For an example, see Figure \ref{fig:sprbase}. Thus, $T$ is changed to
another binary tree on $n$ leaves and we say the two trees are adjacent by an
SPR move.

A nearest neighbor interchange (NNI) move on a tree $T \in \Tr_n$ is determined
by choosing an internal edge $e \in E(T),$ , and rearranging the four subgraphs
(clades) that $e$ induces. It is not difficult to see then that an NNI move is
also  an SPR move.  The following lemma is an application of Proposition
\ref{prop:gengreed} applied to a face of the BME polytope, and two clades
contained in the face.

\begin{lemma}
Let $F $ be a face of the BME polytope $\Po_n$, where $C_1, C_2$ are disjoint
clades with $C_1,C_2\in T, \; \forall \,T\in \vertices(F)$. There exists an
objective $\ve c \in \R^{n \choose 2}$ such that 
\begin{equation*}
\underset{T \in \Tr_n}{\argmax}\, \ve w^T \cdot \ve c = \{\, T \in \vertices (F) \mid d_T(C_1,C_2) \geq d_{T'}(C_1,C_2), \; \forall \, T' \in \vertices(F) \,\}.
\end{equation*}
Similarly there exists an objective $\ve d \in \R^{n \choose 2}$ such that 
\begin{equation*}
\underset{T \in \Tr_n}{\argmax}\, \ve w^T \cdot \ve d = \{\, T \in \vertices (F) \mid d_T(C_1,C_2) \leq d_{T'}(C_1,C_2), \; \forall \, T' \in \vertices(F) \,\}.
\end{equation*}
\label{lem:maxmincladesdist}
\end{lemma}
 For a proof of Lemma \ref{lem:maxmincladesdist} see the appendix. 

\begin{corollary} Every NNI move corresponds to an edge of the BME polytope.
\label{cor:nni}
\end{corollary}

\begin{proof} 
Simply take the objective $\ve c$ given by Lemma \ref{lem:bmeconstcladeface}
which yields the face of the three trees corresponding to an NNI, then add the
extra criteria that either two of the clades are as close or as far as
possible, and apply Lemma \ref{lem:maxmincladesdist}. 
\end{proof} 
\begin{flushright}
$\square$
\end{flushright}

We now present our result that any pair of two trees adjacent by an SPR move yields an edge of the BME polytope.

\begin{theorem}
If $T_1, T_2 \in \Tr_n$ are adjacent by an SPR move, then there exists $\ve c
\in \R^{n \choose 2}$ such that $\ve w^{T_1} \cdot \ve c = \ve w^{T_2} \cdot
\ve c  > \ve w^T \cdot \ve c$ for all $T \in \Tr_n \backslash \{T_1,T_2\}$. \\
\label{thm:spr}
\end{theorem}

\begin{proof}
Let $T_1, T_2 \in \Tr_n$ be adjacent by an SPR move. Any such move can be
described by Figure \ref{fig:sprbase}, where $C_1,\ldots,C_p$ (labeled
$1,\ldots,p$ in Figure \ref{fig:sprbase}) are clades common to $T_1$ and $T_2$
and clade $C_p$ is the subtree that is pruned and regrafted.  By Lemma
\ref{lem:bmeconstcladeface} there exists an objective $\ve c_1 \in \R^{n
\choose 2}$ for the clade-face $F_{C_1,\ldots,C_p}$. Note that $T_1, T_2 \in
F_{C_1,\ldots,C_p}$, but in general $F_{C_1,\ldots,C_p}$ will contain more
trees, hence further restrictions need to be placed on the objective. By
repeated use of Lemma \ref{lem:maxmincladesdist}, and the objective $\ve c_1$
which defines $F_{C_1,\ldots,C_p}$, there exists an objective $\ve c_2 \in
\R^{n \choose 2}$ such that in this order of importance, 
\begin{enumerate}
   \item[1)] the distance between clades $C_{p-1}$ and $C_{p-2}$ is maximized, 
   \item[2)] the distance between clades $C_{1}$ and $C_{2}$ is minimized, 
   \item[3)] the distance between clades $C_{2}$ and $C_{3}$ is minimized, 
   \item[4)] the distance between clades $C_{3}$ and $C_{4}$ is minimized, \\
   $\vdots$ 
   \item[p-4)] the distance between clades $C_{p-4}$ and $C_{p-3}$ is minimized, 
\end{enumerate}
for trees in $F_{C_1,\ldots,C_p}$.  Since $T_1$ and $T_2$ contain the clades
$C_1,\ldots,C_p$ and the properties in the previous list are satisfied in the
prescribed order, we see that $\{T_1,T_2\}\subset \argmax_{T \in \Po_n}\, \ve
w^T \cdot \ve c_2,$ but the latter may also contain the trees with the clades
in the dashed box of Figure \ref{fig:sprbase} inverted vertically.

Select leaves $i \in C_{p-1}$ and $j \in C_1$ such that $d_{T_2}(i,j) \leq
d_{T_2}(m,n)$, $\forall (m,n) \in C_{p-1} \times C_1$.  Similarly, let $k \in
C_{p-2}$ and $l \in C_{p-3}$ such that $d_{T_1}(k,l) \leq d_{T_1}(m,n)$,
$\forall (m,n)\in C_{p-2} \times C_{p-3}$. 

Let $\ve d =  w_{kl}^{T_1} \ve e_{ij} + w_{ij}^{T_2} \ve e_{kl} $. Note that
$2w_{ij}^{T_1} = w_{ij}^{T_2}$ and $w_{kl}^{T_1} = 2 w_{kl}^{T_2}$. It follows
then that $\ve w^{T_1} \cdot \ve d = \ve w^{T_2} \cdot \ve d$. There exists
$\epsilon > 0$ small enough such that \begin{equation*} \min_{\substack{\ve x,
\ve y \in \vertices (\Po)\\ \ve x \in \argmax_{\ve z \in \Po}{\ve c_2 \cdot \ve
z} \\ \ve c_2 \cdot \ve x \neq \ve c_2 \cdot \ve y}}  \ve c_2 \cdot \ve x - \ve
c_2 \cdot \ve y  > \max_{\ve x, \ve y \in \vertices (\Po)} | \epsilon \ve d
\cdot \ve x -  \epsilon \ve d \cdot \ve y |.  \end{equation*} Therefore
Proposition \ref{prop:gengreed} holds and the objective $ \epsilon \ve d$
optimized over $\argmax_{T \in \Tr_n} \ve w^T \cdot \ve c_2$ gives trees such
that either clades $C_1$ and $C_{p-1}$ are as close as possible or $C_{p-2}$
and $C_{p-2}$ are as close as possible. Therefore $\argmax_{T \in \Tr_n} \ve
w^T \cdot (\ve c_2 +\epsilon \ve d)$ contains only $T_1$ and $T_2$.
\begin{flushright}
$\square$
\end{flushright}
\end{proof}
 
\section{Cherry Forcing Objectives}
\label{sec:cfa}
The following lemma is a sufficient condition for vectors $\ve c^1,\ve c^2 \in
\R^{n \choose 2}_+$ to satisfy Proposition \ref{prop:gengreed} on the BME
polytope $\Po_n$.
\begin{lemma}    
\label{lem:gengreedbme}
Let  $\ve c^1, \ve c^2 \in \R_+^{n \choose 2}$  where  for a fixed $K > 0,$
$\ve c^1_{ij} = K$ for all $\{i,j\} \in \supp(\ve c^1).$ 
    If
\begin{equation}
\label{eq:domcondbme}
    \frac{K}{2^{n-2}} > \frac{1}{2} \ve c^2 \cdot (1,1,\ldots,1)
\end{equation}
then 
\begin{equation*}
\underset{T \in \Po_n}{\argmax \,} (\ve c^1 + \ve c^2)\cdot \ve w^T \subseteq \underset{T \in \Po_n}{\argmax \, } \ve c^1 \cdot \ve w^T.
\end{equation*}
\end{lemma}
See the appendix for a proof of Lemma \ref{lem:gengreedbme}.  If a triple $(\ve
c^1, \ve c^2, \Po_n)$ satisfies the assumptions and hypothesis of Lemma
\ref{lem:gengreedbme} and Equation \ref{eq:domcondbme} then we say it satisfies
the \emph{dominance condition}. 
    
Given a clade $C$ in $ \Tr_n$  as input, the idea of the Cherry Forcing
Algorithm is to iteratively fill in entries in an objective $\ve c \in \R^{n
\choose 2}$ to satisfy the dominance condition given in Equation
\ref{eq:domcondbme}  in such a fashion that respects $C.$ More precisely, under
the Cherry Forcing Algorithm, (1) a small part of the topology (e.g. cherry) of
$C$ is fixed, and (2) subsequently filled-in entries in $\ve c$ will be
sufficiently small such that no previously fixed structures of $C$ will be
broken when maximizing $\ve c \cdot \ve w^T$ over $\Po_n$.  It is proved in
Lemma \ref{lem:cfa} that  the sum $\ve {c} = \sum_{i}{\ve c}^i $ of the outputs
of the Cherry Forcing Algorithm will yield the normal vector to the face of the
BME polytope $\Po_n$ that consists of all trees that contain $C$. If $C$ is an
entire tree, then the vector $\ve c$ is in the normal cone of the tree $C$ of
$\Po_n$. That is, $\ve c \cdot \ve w^T$ is maximal only when $C=T$.

\begin{algorithm}[Cherry Forcing Algorithm]
    \mbox{}
    \vskip .25cm
    \label{alg:cfa}
    \begin{center}
                \begin{algorithmic}
                    \INPUT $\widehat T \in \Tr_n$, a clade $\widehat C$ of $\widehat T$. 
                    \OUTPUT $\ve c^1,\ve c^2,\ldots,\ve c^t \in  \R_+^{n \choose 2}$.
                    \STATE Initialization: Let $T_1 := \widehat T$, $K_1 := 1$, $t := 1$, and $\ve c^i := \ve 0 \in  \R^{n \choose 2}$.
                    
                    \REPEAT 
                    \STATE Pick a cherry $\{ k,l\}$ of $T_t$, s.t. the subgraphs of $\widehat T$ given by $k$ and $l$ are in clade $\widehat C$.
                    
                        \STATE Let $G_k$ be the nodes of the subgraph of $\widehat T$ given by $k$.
                        \STATE Let $G_l$ be the nodes of the subgraph of $\widehat T$ given by $l$.
                        \FOR{every pair $\{p,q\} \in G_k \, \times \, G_l$}
                            \STATE Let $c^t_{p,q} := \frac{ K_t}{{n \choose 2}}$.
                        \ENDFOR

                        \STATE Let $ K_{t+1} := \frac{ K_t}{{n \choose 2}2^{n-1}} $.
                        \STATE Let $t := t + 1$.
                        \STATE Let $T_t :=$ $T_{t-1}$ where cherry $\{k,l\}$ is amalgamated.
                        
                    \UNTIL $T'$ has a single leaf corresponding to the entire clade $\widehat C$ of $\widehat T$ or $T'$ is the star tree on three leaves.
                    \RETURN  $\ve c^1,\ve c^2,\ldots,\ve c^t$.
                \end{algorithmic}
    \end{center}
\end{algorithm}

\begin{lemma}
Let $\widehat T \in \Tr_n$ and clade $\widehat C$ of $\widehat T$ be the input
of Algorithm \ref{alg:cfa} with output $\ve c^1, \ve c^2,\ldots, \ve c^t$.
Every triple 
\label{lem:cfadomcond}
\begin{align*}
    &\{\ve c^1, \ve c^2 + \cdots + \ve c^t, \Po_n \}, \\
    &\{\ve c^2, \ve c^3 + \cdots + \ve c^t, \Po_n \}, \\
    &\{\ve c^3, \ve c^4 + \cdots + \ve c^t, \Po_n \}, \\
    &\phantom{AAAA} \vdots \\
    &\{\ve c^{t-1}, \ve c^t, \Po_n \}
\end{align*}
    satisfies the dominance condition in Equation \ref{eq:domcondbme}. Consequently $\argmax_{T \in \Tr_n}  (\sum_{i=1}^t \ve c^i) \cdot  \ve w^T = \{\, T \in \Tr_n \mid \widehat C \text{ is a clade of } T\, \}$.
    
\label{lem:cfa}    
\end{lemma}
 A proof of Lemma \ref{lem:cfa} is provided  in the appendix.

\begin{lemma} $\phantom{A}$ 
\label{lem:cfadisjoint}
Let $\widehat T \in \Tr_n$ and clade $\widehat C$ of $\widehat T$ be the input
of Algorithm \ref{alg:cfa} with output $\ve c^1, \ve c^2,\ldots, \ve c^t$.
Then $\supp(\ve c^i) \, \cap \, \supp(\ve c^j) = \emptyset$ for all $1 \leq i <
j \leq t$.
\end{lemma}
 A proof of Lemma \ref{lem:cfadisjoint} is provided  in the appendix.

\section{Non-empty Intersection of NJ and BME Cones}
\label{sec:njbme}
The NJ Algorithm, first presented in \cite{Saitou1987}, is a consistent
distance-based method to reconstruct a phylogenetic tree. Yet, its biological
interpretation and what criteria it optimized have only been established
recently. Some initially argued that NJ optimized an ordinary least-squares
criteria at each step, while others contended that it did not optimize any
criteria.  See \cite{Steel2006} for a short history of NJ. However in
\cite{Steel2006}, it was shown that in fact, NJ greedily minimizes the BME
criteria at every neighbor joining step. In \cite{kord2009} Eickmeyer et.  al.
characterized those dissimilarity maps for which the output of the NJ Algorithm
is in fact  the BME tree, by a comparison of the NJ cones with the BME cones,
for eight or fewer taxa.

Given a tree topology $T \in \Tr_n$ with branch lengths $\omega$, it follows
from consistency that NJ and BME will return $T$ given the tree metric
$D_{(T,\omega)}$ defined in Section 2.1.  That is, $D_{(T,\omega)}$ will lie in at
least one NJ cone of $T$.  The order in which NJ picks cherries depends on the
dissimilarity map, and the dissimilarity map depends on the branch lengths.
Therefore which NJ cone $D_{(T,\omega)}$ lies in is strictly determined by the
branch lengths $\omega$.  However, if a NJ cone $C$ of $T$ is fixed
it is not clear how branch lengths, call it $\omega'$, can be assigned to the
tree topology of $T$ such that $D_{(T,\omega')}$ is in $C$. Thus, it is not
clear that consistency directly implies that the BME cone and every NJ cone
have non-trivial intersection.

Our result is that every NJ cone associated to a tree topology $T \in \Tr_n$
has an intersection of positive measure with the BME cone. That is, for any NJ
cone associated with the particular order to pick cherries and the tree $T$,
there is an intersection of positive measure with the BME cone associated to
$T$, where the BME cone is defined as the set of all dissimilarity maps $\ve d
\in \R^{n \choose 2}$ such that $\argmax_{T' \in \Tr_n} \ve d \cdot \ve w^{T'}
\supseteq \{T\}$.

The NJ Algorithm takes as input a dissimilarity map $\ve c \in \R^{n \choose
2}$ and builds a tree $T \in \Tr_n$ \cite{Saitou1987}.  It involves: 1) picking
a cherry $\{i,j\}$, 2) creating a node $a$ joining taxa $i$ and $j$, 3)
computing the distances from other nodes to the new node $a$, 4) repeating the
procedure until the number of leaves $n$ is 3. 

The main problem is picking the cherry. A solution, suggested by  Saitou and
Nei \cite{Saitou1987} and subsequently modified by Studier and Keppler
\cite{Studier1988}, relies on the $Q$-criterion in Theorem \ref{simplecherry}
below.

\begin{theorem}[Cherry-picking criterion ($Q$-criteria) \cite{Saitou1987, Studier1988}]
\label{simplecherry}
Let $\ve c \in \R^{n \choose 2}$ be an additive tree metric for a tree $T\in
\Tr_n$ and define the $n\times n$-matrix $Q_{\ve c}$ with entries:
\begin{equation}
 Q_{\ve c}(i,j) = (n-2)c_{i,j} - \sum_{k=1}^n c_{i,k} - \sum_{k=1}^n c_{k,j} = (n-4)c_{i,j} - \sum_{k\neq j} c_{i,k} - \sum_{k\neq i} c_{k,j}.
\label{eq:qcriteria}
\end{equation}
Then any pair of leaves $\{i^*,j^*\}$, for which $Q_{\ve c}(i^*, j^*)$  is
minimal, is a cherry in the tree $T$.
\end{theorem}

If the NJ Algorithm selects taxa $\{k,l\}$ as a cherry, and $a$ is the new node
joining $\{k,l\}$ then the new dissimilarity map $\ve c' \in \R^{n-1 \choose
2}$ is defined to be
\begin{align*}
\text{if } i \neq a \neq j \qquad & c'_{i,j} = c_{i,j}\\
\text{else } \qquad & c'_{i,a} = \frac{1}{2}\Big( c_{i,k} + c_{i,l} - c_{k,l} \Big).
\end{align*}

\begin{lemma}[Shifting Lemma \cite{Eickmeyer2008}]
Let $\ve c ,\ve x \in \R^{n \choose 2}$ where $\ve x = (1,1,\ldots,1)$. Then
the Neighbor Joining Algorithm applied to $\ve c + k\ve x$, for any $k \in \R$,
returns the same tree as the Neighbor Joining Algorithm applied to $\ve c$.
Moreover, the linear ordering of the $Q$-criteria of $\ve c$ is the same as the
linear ordering of the $Q$-criteria of $\ve c + k\ve x$, i.e. if $Q_{\ve
c}(i_1,j_1) \leq(<)\, Q_{\ve c}(i_2,j_2)$ then $Q_{\ve c+k \ve x}(i_1,j_1)
\leq(<)\, Q_{\ve c+k \ve x}(i_2,j_2)$.
\label{lem:shifting}
\end{lemma}

\begin{lemma}

Let $\ve c^1, \ve c^2 \in \R_+^{n \choose 2}$ where $|\supp(\ve c^1)| = 1$, and
$(\ve c^1, \ve c^2, \Po)$ satisfies the dominance condition (Equation
\ref{eq:domcondbme}).  Further, let $\ve c = \ve c^1 + \ve c^2$, $\{\{p,q\}\} =
\supp(\ve c^1)$, and $Q_{- \ve c}$ be the Q-criteria calculated from the
dissimilarity map $- \ve c$. 

If $n > 4$, $Q_{- \ve c}(p,q) < Q_{- \ve c}(i,j)$ for all $\{i,j\} \neq
\{p,q\}$. If $n=4$, and $p,q,r,s$ are the leaves, then $Q_{- \ve c}(p,q) = Q_{-
\ve c}(r,s) < Q_{- \ve c}(i,j)$ where $\{i,j\} \neq \{p,q\}$ and $\{i,j\} \neq
\{r,s\}$.
\label{lem:expcQ}
\end{lemma}
 See the appendix for a proof of Lemma \ref{lem:expcQ}.
\begin{theorem}
\label{thm:njbme}
Let $\widehat T \in \Tr_n$ be the input for Algorithm \ref{alg:cfa} with output
$\{\ve c^1,\ldots,\ve c^t\}$.  The Neighbor Joining Algorithm with input $-(\ve
c^1 + \cdots + \ve c^t)$ returns $\widehat T$.
\end{theorem}
\begin{proof}
 
We proceed by induction on $n$. If $n=3$ we have the star tree and there is
nothing to be done. If $n=4$ then Lemma \ref{lem:cfa} applies to the output of
Algorithm \ref{alg:cfa} and we are done since NJ will return $\widehat T$ by
Lemma \ref{lem:expcQ}. Consider $n > 4$.  Let $\widehat T \in \Tr_n$ be the
input for Algorithm \ref{alg:cfa} with output $\{\ve c^1,\ldots,\ve c^t\}$.
Define $\ve c = \ve c^1 + \cdots + \ve c^t$. Note also that by Lemma
\ref{lem:cfadisjoint}, $\supp(\ve c^1), \ldots, \supp(\ve c^t)$ are pairwise
disjoint.  We know Lemma \ref{lem:cfadomcond}  applies and implies by Lemma
\ref{lem:gengreedbme},  $\argmax_{T \in \Tr_n} \ve w^T \cdot \ve c = \{\widehat
T\}$.  Let $\{\{p,q\}\} = \supp(\ve c^1)$ be the first cherry picked in
Algorithm \ref{alg:cfa} and let $Q_{-\ve c}$ be the $Q$-criteria of $-\ve c$.
By Lemma \ref{lem:expcQ},  $Q_{-\ve c}(p,q)$ is the minimal element in $Q_{-\ve
c}$. Consider the shifted vector $\ve d := \ve 1 - {n \choose 2}\ve c \in \R^{n
\choose 2}$, and note that $d_{p,q} = 0$.  The Shifting Lemma
\ref{lem:shifting} implies that $Q_{\ve d}(p,q)$ will be the minimal element in
$Q_{\ve d}$.  Thus, the NJ Algorithm will join leaves $p$ and $q$ to the new
node $a$. Consider the new dissimilarity map $\ve d' \in \R^{n-1 \choose 2}$
given by the NJ Algorithm. If $i \neq a$ and $j \neq a$ then $d'_{i,j} =
d_{i,j}$. For $i \neq a$, $d'_{i,a} = \frac{1}{2}( d_{i,p} + d_{i,q} - d_{p,q})
= \frac{1}{2}( d_{i,p} + d_{i,q})$, since $d_{p,q} = 0$.  Since the cherry
$\{p,q\}$ was designated first by Algorithm \ref{alg:cfa}, by construction, for
all $i \neq p$ and $i \neq q$, $c^l_{i,p} = c^l_{i,q}$ for all $1 < l \leq t$.
This implies $d'_{i,a} = d_{i,p} = d_{i,q}$ for $i \neq a$.  Define $\ve
{\widehat c} := \ve 1 - \ve d'$ and  $\ve {\widehat c}^2,\ldots,\ve {\widehat
c}^t \in \R_+^{n-1 \choose 2}$ as follows \begin{equation*} \widehat c_{i,j}^l
= \left\{ \begin{array}{ll} \widehat c_{i,j} & \text{   if } \{i,j\} \in
\supp(\ve c^l) \\ 0 & \text{else} \end{array} \right.  \end{equation*} for $2
\leq l \leq t$.  Observe that $|\supp(\ve {\widehat c}^2)| = 1$ since $\ve
{\widehat c}^2$ corresponds to the  cherry picked in the amalgamated tree $T_2$
in Algorithm \ref{alg:cfa}, and $p$ and $q$ have been identified with $a$.
Moreover, every triple $(\ve {\widehat c}^2, \ve {\widehat c}^3 + \cdots + \ve
{\widehat c}^t, \Po_{n-1})$, $(\ve {\widehat c}^3, \ve {\widehat c}^4 + \cdots
+ \ve {\widehat c}^t, \Po_{n-1})$, $\ldots$, $(\ve {\widehat c}^{t-1},  \ve
{\widehat c}^t, \Po_{n-1})$ satisfies the dominance condition of Equation
\ref{eq:domcondbme}. Thus $\argmax_{T \in \Tr_n} \ve {\widehat c} \cdot \ve w^T
= \widehat T'$ for some $\widehat T' \in \Tr_n$.  Since $\ve {\widehat c}$ is a
dissimilarity map on $n-1$ leaves, the induction hypothesis holds, and NJ
returns $\widehat T'$.  Finally $\widehat T'$ is contained in $\widehat T$ as a
clade, which implies NJ on $-\ve c$ will return $\widehat T$, since $\widehat
T$ equals the tree $\widehat T'$ with leaves $i$ and $j$ connected to leaf $a$
by two different edges.

\begin{flushright}
$\square$
\end{flushright}
\end{proof}

Given a fixed tree topology, Algorithm \ref{alg:cfa} allows for any choice of
neighbor joining pairs (cherries in the NJ Algorithm), and every such choice
yields a different NJ cone. Thus, Theorem \ref{thm:njbme} implies that every NJ
cone and BME cone have a non-empty intersection.

\begin{corollary} 
Every NJ cone $C$ associated to a fixed $T \in \Tr_n$ has an intersection of
positive measure with the BME cone associated to $T$.
\end{corollary}
\begin{proof}
Let $T \in \Tr_n$ be a tree topology and $C$ be a NJ cone associated to $T$;
recall $C$ is also dependent upon an order of picking cherries.  Now apply
Algorithm \ref{alg:cfa} with $T$ as the input (as both the tree and clade),
choosing cherries in step 5 by the order associated to the NJ cone $C$, and let
$\{\ve c^1,\ldots, \ve c^t\}$ be the output.   By Theorem \ref{thm:njbme}, the
BME and NJ algorithm with input $-\sum_{i=1}^t \ve c^i$ will each return $T$.
Moreover, since the cherries were chosen in step 5 to be consistent with $C$,
we have $-\sum_{i=1}^t \ve c^i \in C$. \\ \indent Since the BME cone associated
to $T$ is convex (as a normal cone of the BME polytope), and $\argmin_{T' \in
\Tr_n} (-\sum_{i=1}^t \ve c^i) = \{T \}$ by Lemma \ref{lem:cfa}, it follows
that $-\sum_{i=1}^t \ve c^i$ lies in the interior of the BME cone associated to
$T$.\\ \indent On the other hand, individual NJ cones are convex (by
definition, or see \cite{kord2009}) and the boundary of the intersection of
multiple NJ cones associated to the same tree topology corresponds to two or
more cherries having equal Q-scores (i.e., $Q$-criteria entries) at some step
in the NJ Algorithm \cite{kord2009}. Lemma \ref{lem:expcQ} implies that  the
first cherry chosen by the NJ Algorithm will be $\supp(\ve c^1)$, that is, it
has the smallest Q-score with no ties. Moreover, in the proof of Theorem
\ref{thm:njbme} we see that the new dissimilarity map derived from
$-\sum_{i=1}^t \ve c^i$ in the NJ Algorithm also satisfies the dominance
condition. Hence, Lemma \ref{lem:expcQ} holds again, and there are no ties in
the Q-score. Therefore, there will be no ties in the Q-score, except for the
case of four taxa. For four taxa, the only ties present are the trivial ones:
If $S$ is the set of four taxa ($|S| = 4$) then by definition of the Q-score, 
\begin{equation*}
Q(p,q) = Q(r,s) \qquad \forall\; \{p,q\} \subseteq S,\, \{r,s\} = S \setminus \{p,q\}.
\end{equation*}
We note that these trivial ties do not correspond to different NJ cones, and
hence $-\sum_{i=1}^t \ve c^i$ lies in the interior of C. \\ \indent In
conclusion, we see that  $-\sum_{i=1}^t \ve c^i$ lies in the  interiors of both
the BME cone and the NJ cone $C$. This implies they have an intersection of
positive measure.
\begin{flushright}
$\square$
\end{flushright}
\end{proof}

\section{Discussion}
\label{sec:disc}
Mathematically, ``closeness'' between trees is measured via differing distances
(metrics) on tree space $\mathcal{T}_n, $ including the popular distance
measures $d_{NNI}(T,T'),$ $d_{SPR}(T,T'),$ and $d_{TBR}(T,T')$ describing the
minimum number of nearest neighbor interchange (NNI) (resp.,
subtree-prune-regrafting (SPR), tree-bisection-regrafting (TBR)) moves needed
to transform $T$ to $T'$ for $T,T'\in \mathcal{T}_n.$ Each such metric $M$
yields a notion of adjacency, with $T,T'\in \mathcal{T}_n$ being $M$ adjacent
if $d_{M}(T,T') = 1.$ The comparisons of two trees $T,T'$ as NNI, SPR, or TBR
adjacent confers useful biological information, including providing the basis
for multiple tree reconstruction algorithms
\cite{Bonet2009,Steel1993,Semple2003,Steel2004}. For $T,T'\in \mathcal{T}_n$ ,
set  $d_{BME}(T,T') = 1$  if  $\mathbf{w}^T$ and $\mathbf{w}^{T'}$ are two
vertices joined by an edge in the BME polytope $\mathcal{P}_n. $ This yields
another notion of adjacency in the BME setting.    

The point of view of this paper is that knowledge of BME adjacency, and its
relationship to NJ adjacency, has likewise the potential to inform our
understanding of tree space $\mathcal{T}_n,$ and the gene and/or species trees
its elements represent. We have explored some relationships between adjacency
for $M = $ NNI, SPR, TBR, BME, and NJ.  It is well-known that an NNI move is a
special case of an SPR move and an SPR move is a special case of a TBR move.
In this paper, we have shown that SPR adjacency implies BME adjacency. However
it is not known that TBR adjacency implies BME adjacency.  We have made some
initial explorations in this regard, including using an additional related
notion of ``circular adjacency'' predicated upon the circular orderings
employed in \cite{Steel2004}. However, having seen no examples to show that
TBR adjacency fails to imply BME adjacency, we propose the following
conjecture.
\begin{conjecture}\label{con1} 
If $T,T'\in \mathcal{T}_n,$ then $d_{TBR}(T,T') = 1$ implies $d_{BME}(T,T') =
1.$ 
\end{conjecture}
Considering further the potential applications of such adjacency notions in the
context of the BME polytopes and BME cones is a topic we hope to explore in a
future work.

\section{Acknowledgments}
D. Haws and R. Yoshida are supported by NIH R01 grant 5R01GM086888. T. Hodge is
supported by NSF (DUE) grant 0737467.  The authors would like to thank P.
Huggins for all his helpful discussions, C. Segroves for directing us to a
useful reference, and M. Cueto for pointing out a typo in the manuscript. The
authors would also like to thank the referees for detailed critiques and
multiple suggestions for improving the paper.

\bibliographystyle{amnat2}

\bibliography{reference}

\appendix

\section{Appendix}

\begin{proof}[Lemma \ref{lem:bmeconstcladeface}]

The proof of the lemma relies almost entirely on the proof in \cite{BME} of the
consistency of the BME method for phylogenetic tree reconstruction
(\cite[Theorem 2 and Appendix 3]{BME}]).  Given $T\in \Tr_n$ with
edge-weighting $\omega,$ recall the notation of Section 2.1 and 2.3.
Furthermore, from Equations (\ref{e:DTomega}) and (\ref{eq:pauplin}), for any
tree $T'\in \Tr_n$ one can obtain the estimated BME length of $T'$ as a linear
function of the metric $D:=D_{T,\omega}$ as
\begin{equation}\label{eq:linbmelength}
 l(T',D) = \ve w^{T'} \cdot D = \sum_{A\,|\,B \in \Sigma(T)} \omega(A\,|\,B) l(T',D^{A\,|\,B})
\end{equation}  
 
By the consistency of the BME tree length estimation, $l(T,D) = l(T)
=\omega(T).$ So, for the proof of the consistency of the BME method it sufficed
for \cite{BME} to demonstrate that $ l(W,D) >  l(T,D)$ for all $W\in \Tr_n$
with $W\not= T.$  By Equation (\ref{eq:linbmelength}), it was enough to prove
this inequality holds for any split metric $D^{A\,|\,B}$ of $T$ in place of
$D.$ Likewise, for our proof of Lemma \ref{lem:bmeconstcladeface}, we consider
the tree $T \in \Tr_n$ such that $C_1,\ldots,C_p \in T$. Furthermore, we take
an edge-weighting $\omega$  of $T$ for which $\omega(e) = 0$ if $e\notin C_i,$
for all $1\leq i \leq p.$  We will show that if $W \in \Tr_n$ contains
$C_1,\ldots,C_p$ then $ l(W,D) =  l(T,D)$. Otherwise if $C_i\notin W$ for some
$1\leq i\leq p,$  then we show $l(W,D) > l(T,D).$ Both parts proceed by
reducing to the case of split metrics $D^{A\,|\,B}$ for $T$, and drawing upon
the results in \cite{BME}.

Consider $W \in \Tr_n$ such that $ C_1,\ldots,C_p \in W$. Since $C_i\in W,T$
for all $1\leq i\leq p,$ $\Sigma(W) \cap \Sigma(T)$ contains any split
$A\,|\,B$ induced by any edge $e\in C_1,\ldots, C_p.$ As shown in \cite{BME}
(by direct calculations using the definition of the BME branch lengths $l$), if
a split $A\,|\,B$ is both in $\Sigma(T)$ and $\Sigma(W)$ then $l(W,D^{A\,|\,B})
= l(T,D^{A\,|\,B}) = 1$.  If $W \neq T$ then there exist some split $A\,|\,B$
in $\Sigma(T)$ but not in $\Sigma(W),$ and  $\omega(A\,|\,B) = 0$. Thus $l(W,D)
=  l(T,D)$. Now consider $W \in \Tr_n$ such that $C_i\notin W$ for some $1\leq
i\leq p.$ As above, there exists some split $A\,|\,B$ in $\Sigma(T)$ and not in
$\Sigma(W)$, and moreover $\omega(A\,|\,B) > 0$. Under these circumstances, the
argument in the (remainder of the) proof of Theorem $2$ of \cite{BME} applies
to show $ l(W,D^{A\,|\,B}) > l(T,D^{A\,|\,B})$, which suffices to complete the
proof of Lemma \ref{lem:bmeconstcladeface}. 

\begin{flushright}
$\square$
\end{flushright}
\end{proof}
\begin{proof}[Proposition \ref{prop:gengreed}]
Let $\ve a, \ve b \in \vertices(\Po)$ and suppose $\ve a \in \underset{\ve z
\in \vertices(\Po)}{\argmax} \, \ve z \cdot \ve c_1 \not \ni \ve b$. Thus $\ve
a \cdot \ve c_1 > \ve b \cdot \ve c_1$. Then
\begin{align*}
& \ve a \cdot (\ve c_1 + \ve c_2) - \ve b \cdot (\ve c_1 + \ve c_2)  = \ve a \cdot \ve c_1 - \ve b \cdot \ve c_1 + \ve a \cdot \ve c_2 - \ve b \cdot \ve c_2\\
& \geq \min_{\substack{\ve x, \ve y \in \vertices (\Po)\\ \ve x \in \argmax_{\ve z \in \Po}{\ve c_1 \cdot \ve z} \\ \ve c_1 \cdot \ve x \neq \ve c_1 \cdot \ve y}}  \ve c_1 \cdot \ve x - \ve c_1 \cdot \ve y  + \ve a \cdot \ve c_2 - \ve b \cdot \ve c_2 > 0,
\end{align*}
since 
\begin{equation*}
\min_{\substack{\ve x, \ve y \in \vertices (\Po)\\ \ve x \in \argmax_{\ve z \in \Po}{\ve c_1 \cdot \ve z} \\ \ve c_1 \cdot \ve x \neq \ve c_1 \cdot \ve y}}  \ve c_1 \cdot \ve x - \ve c_1 \cdot \ve y  > \max_{\ve x, \ve y \in \vertices (\Po)} | \ve c_2 \cdot \ve x - \ve c_2 \cdot \ve y |.
\end{equation*}
Thus $\ve b \notin \underset{\ve z \in \vertices (\Po)}{\argmax}\, (\ve c_1 +
\ve c_2) \cdot \ve z$ and therefore 
\begin{equation*}
\underset{\ve z \in \vertices (\Po)}{\argmax}\, (\ve c_1 + \ve c_2) \cdot \ve z \subseteq \underset{\ve z \in \vertices (\Po)}{\argmax}\, \ve c_1 \cdot \ve z.
\end{equation*}
\begin{flushright}
$\square$
\end{flushright}
\end{proof}

\begin{proof}[Lemma \ref{lem:maxmincladesdist}]
Let $\ve d \in \R^{n \choose 2}$ be the normal vector of the face $F$. Let $i$
be a leaf of clade $C_1$ and $j$ a leaf of clade $C_2$. Now apply Proposition
\ref{prop:gengreed} as follows: There exists $\epsilon > 0$ sufficiently small
such that 
\begin{equation*}
\min_{\substack{\ve x, \ve y \in \vertices (\Po_n)\\ \ve x \in \argmax_{\ve z \in \Po_n} \ve z \cdot \ve d\\ \ve d \cdot \ve x \neq \ve d \cdot \ve y}}  \ve d \cdot \ve x - \ve d \cdot \ve y  > \max_{\ve x, \ve y \in \vertices (\Po_n)} | \epsilon (-\ve e_{ij})\cdot \ve x - \epsilon (- \ve e_{ij}) \cdot \ve y |.
\end{equation*}
Therefore 
\begin{equation*}
\underset{T \in \Tr_n}{\argmax} \, \ve w^T \cdot (\ve d - \epsilon \ve e_{ij}) = \underset{T \in \vertices(F)}{\argmax}\, \ve w^T \cdot (- \epsilon \ve e_{ij})
\end{equation*}
which are precisely all trees contained in $\vertices(F)$ such that clades
$C_1$, $C_2$ are farthest apart. To show there exists an objective
corresponding to a face contained in $\vertices(F)$ such that clades $C_1$,
$C_2$ are close as possible, simply change $-\ve e_{ij}$ to $\ve e_{ij}$. 
\begin{flushright}
$\square$
\end{flushright}
\end{proof}

\begin{proof}[Lemma \ref{lem:gengreedbme}]
Consider the left-hand side of the inequality in Equation \ref{eq:gengreed}
applied to $\ve c^1, \ve c^2$ and $\Po_n:$
\begin{equation}         
    \min_{\substack{ T'',T' \in \Po_n \\ T'' \in \argmax_{T \in \Tr_n } \ve c^1 \cdot \ve w^T \\ \ve c^1 \cdot \ve w^{T''} > \ve c^1 \cdot \ve w^{T'}  }}  K \left( \sum_{\{i,j\} \in \supp(\ve c^1)}  (w_{ij}^{T''} - w_{ij}^{T'}) \right).
\label{eq:gengreedbme:lhs}
\end{equation}         
First note that $ \frac{1}{2^{n-2}} \leq w^T_{ij} \leq  \frac{1}{2}$ for all $T
\in \Tr_n$ and all $\{i,j\}$.
Thus, 
\begin{equation*}
2^{n-2} \sum_{\{i,j\} \in \supp(\ve c^1)}  (w_{ij}^{T''} - w_{ij}^{T'})
\end{equation*}
will be integral and greater than $0$. This implies that  
\begin{equation*}
\sum_{\{i,j\} \in \supp(\ve c^1)}  (w_{ij}^{T''} - w_{ij}^{T'}) \geq \frac{1}{2^{n-2}}.
\end{equation*}
Hence the expression in Equation \ref{eq:gengreedbme:lhs} will be greater than
or equal to $\frac{K}{2^{n-2}}$. Using the bounds on $w^T_{ij}$ for all $T$ and
$\{i,j\},$ and the triangle inequality,  
\begin{equation*}
\frac{1}{2} \ve c^2 \cdot (1,1,\ldots,1) \geq \max_{T',T'' \in \Tr_n} | \ve c_2 \cdot \ve w^{T''} - \ve c_2 \cdot \ve w^{T'} |.
\end{equation*}
Therefore if Equation \ref{eq:domcondbme} holds then Proposition
\ref{prop:gengreed} holds, completing the proof of Lemma \ref{lem:gengreedbme}.
\begin{flushright}
$\square$
\end{flushright}
\end{proof}

\begin{proof}[Lemma \ref{lem:cfa}]
Let $\widehat T \in \Tr_n$ and clade $\widehat C$ of $\widehat T$ be the input
of Algorithm \ref{alg:cfa} with output $\ve c^1, \ve c^2, \ldots, \ve c^t$.
Moreover, let $K_1,K_2,\ldots,K_t$ be the ordered list of $K_i$ used.  Let $ 1
\leq r < t$ and from Step 10 in Algorithm \ref{alg:cfa} we see that $K_{r+l} <
\frac{1}{2^l}\frac{1}{2^{n-2}}\frac{K_r}{{n \choose 2}}$ for $1 \leq l \leq
t-r$.
    
Then
\begin{align*}
& \frac{1}{2}\left( \ve c^{r+1} + \ve c^{r+2} + \cdots + \ve c^t \right) \cdot (1,1,\ldots,1) \\
= & \frac{1}{2} \left( \sum_{\{i,j\} \in \supp(\ve c^{r+1})} c^{r+1}_{i,j} + \sum_{\{i,j\} \in  \supp(\ve c^{r+2})} c^{r+2}_{i,j} + \cdots + \sum_{\{i,j\} \in  \supp(\ve c^{t})} c^{t}_{i,j} \right) \\
= & \frac{1}{2} \left( \sum_{\{i,j\} \in  \supp(\ve c^{r+1})} \frac{K_{r+1}}{{n \choose 2}} + \sum_{\{i,j\} \in  \supp(\ve c^{r+2})} \frac{K_{r+2}}{{n \choose 2}} + \cdots + \sum_{\{i,j\} \in  \supp(\ve c^{t})} \frac{K_{t}}{{n \choose 2}} \right) \\
 < &\frac{1}{2}\left( K_{r+1} + \cdots + K_t \right) \\
 < &\frac{1}{2}\left( \frac{1}{2}\frac{1}{2^{n-2}}\frac{K_{r}}{{n \choose 2}} +  \frac{1}{2^2}\frac{1}{2^{n-2}}\frac{K_{r}}{{n \choose 2}} + \cdots + \frac{1}{2^{t-r}}\frac{1}{2^{n-2}}\frac{K_{r}}{{n \choose 2}}  \right) \\
 = &\frac{1}{2}\frac{1}{2^{n-2}}\frac{K_{r}}{{n \choose 2}}\left( \frac{1}{2} +  \frac{1}{2^2} + \cdots + \frac{1}{2^{t-r}}  \right) < \frac{K_r}{{n \choose 2}2^{n-2}} = \frac{c^r_{ij}}{2^{n-2}}
\end{align*}
for $\{i,j\} \in \supp(\ve c^r)$. Thus $(\ve c^r, \ve c^{r+1} + \cdots + \ve
c^{t}, \Po_n)$ satisfies the dominance condition for $ 1 \leq r < t$ and Lemma
\ref{lem:gengreedbme} applies. Therefore $\argmax_{T \in \Tr_n} (\ve c^{r+1} +
\cdots + \ve c^t)\cdot \ve w^T \subseteq \argmax_{T \in \Tr} \ve c^r \cdot \ve
w^T$. Altogether this implies that $\argmax_{T \in \Tr_n} (\ve c^{1} + \cdots +
\ve c^t)\cdot \ve w^T$ is the set of trees where, and in this order, $\ve c^1
\cdot \ve w^T$ is maximized, $\ve c^2 \cdot \ve w^T$ is maximized, $\ldots$,
$\ve c^t \cdot \ve w^T$ is maximized. But this recursive linear optimization of
$\ve c^1, \ldots, \ve c^t$ over $\Po_n$ precisely forces the amalgamation of
cherries determined in Algorithm \ref{alg:cfa}.
\begin{flushright}
$\square$
\end{flushright}
\end{proof}

\begin{proof}[Lemma \ref{lem:cfadisjoint}] \\
Suppose not. That is, for some $1 \leq i < j \leq t$ and some $ 1 \leq k \leq l
\leq n$, assume $c^i_{kl} > 0$ and $c^j_{kl} > 0$. Since $i < j$ this implies
leaves $k$ and $l$ are contained in two separate leaves of $T_i$ in Algorithm
\ref{alg:cfa}. Moreover since $c^i_{kl} > 0$, this implies the leaves of $T_i$
that contained $k$ and $l$ were amalgamated, giving $T_{i+1}$. Thus, leaves $k$
and $l$ will never appear in separate leaves of $T_r$ for any $r > i$. But,
$c^j_{kl} > 0$, implying $k$ and $l$ appear in separate leaves of $T_j$, a
contradiction. \\\\
\begin{flushright}
$\square$
\end{flushright}
\end{proof}

\begin{proof}[ Lemma \ref{lem:expcQ}]
Let $\ve c^1, \ve c^2 \in \R_+^{n \choose 2}$ where $|\supp(\ve c^1)| = 1$, and
$(\ve c^1, \ve c^2, \Po)$ satisfies the dominance condition (Equation
\ref{eq:domcondbme}).  Further, let $\ve c = \ve c^1 + \ve c^2$, $\{p,q\} =
\supp(\ve c^1)$, and $Q_{- \ve c}(i,j)$ be the Q-criteria calculated from the
dissimilarity map $- \ve c$.  If $n =4$, and $p,q,r,s$ are the leaves, then we
see directly that $Q_{- \ve c}(p,q) = Q_{- \ve c}(r,s)$.  Moreover $c^1_{p,q}$
appears in $Q_{- \ve c}(p,r), Q_{- \ve c}(q,s), Q_{- \ve c}(p,s), Q_{- \ve
c}(q,r)$, and not in $Q_{- \ve c}(p,q)$ and $Q_{- \ve c}(r,s)$. The dominance
condition implies $Q_{- \ve c}(p,q) = Q_{- \ve c}(r,s) < \min \Big( Q_{- \ve
c}(p,r), Q_{- \ve c}(q,s), Q_{- \ve c}(p,s), Q_{- \ve c}(q,r) \Big)$.

Now consider $n > 4$.  Since $(\ve c^1, \ve c^2, \Po)$ satisfies the dominance
condition, it follows that
\begin{align*}
\frac{c^1_{p,q}}{2^{n-2}} = \frac{c_{p,q}}{2^{n-2}} &> \frac{1}{2} \left( \sum_{1 \leq i < j \leq n} c^2_{i,j} \right) = \frac{1}{2} \left( \sum_{\{i,j\} \neq \{p,q\}} c_{i,j} \right) \quad &\Rightarrow\\
 \frac{1}{2}c_{p,q} & > \sum_{\{i,j\} \neq \{p,q\}} c_{i,j} \quad &\Leftrightarrow\\
 -\frac{1}{2}c_{p,q} & > - c_{p,q} + \sum_{\{i,j\} \neq \{p,q\}} c_{i,j}.
\end{align*}
This implies
\begin{equation*}
Q_{- \ve c}(p,q) = -(n-4)c_{p,q} + \sum_{k\neq q} c_{p,k} + \sum_{k\neq p} c_{k,q} < -(n-5)c_{p,q} - \frac{1}{2}c_{p,q}.
\end{equation*}
Furthermore, $\frac{1}{2}c_{p,q} > c_{i,j}$ for all $\{i,j\} \neq \{p,q\}$,
since the dominance condition is satisfied. Finally 
\begin{equation*}
Q_{- \ve c}(p,q) < -(n-5)c_{p,q} - \frac{1}{2}c_{p,q} < -(n-5)c_{i,j} - c_{i,j} = -(n-4)c_{i,j} \leq Q_{- \ve c}(i,j).
\end{equation*}
\begin{flushright}
$\square$
\end{flushright}
\end{proof}

\end{document}